\documentclass[fleqn,twoside]{article}
\usepackage{espcrc2}
\usepackage{graphicx}
\usepackage[figuresright]{rotating}

\newcommand{\lesssim}{\mathbin{\lower 3pt\hbox
   {$\rlap{\raise 5pt\hbox{$\char'074$}}\mathchar"7218$}}} %< or of order
\newcommand{\gtrsim}{\mathbin{\lower 3pt\hbox
   {$\rlap{\raise 5pt\hbox{$\char'076$}}\mathchar"7218$}}} %> or of order
\newcommand{\Msun}{M_{\odot}}
\newcommand{\Mearth}{M_{\oplus}}
\newcommand{\be}{\begin{eqnarray}}
\newcommand{\ee}{\end{eqnarray}}
\newcommand{\DM}{\rm DM}

\newcommand{\SM}{\rm SM}
\newcommand{\RM}{\rm RM}
\newcommand{\nel}{n_{\rm e}}
\newcommand{\cnsq}{C_{\rm n}^2}
\newcommand{\Bpar}{B_{\parallel}}
\newcommand{\Pdot}{\dot P}
\newcommand{\Bq}{B_{\rm q}}

\newcommand{\Sminone}{S_{\rm min_1}}

\newcommand{\Lp}{L_{\rm p}}

\newcommand{\Tsys}{T_{\rm sys}}
\newcommand{\Ssys}{S_{\rm sys}}

\newcommand{\Dmax}{D_{\rm max}}

\newcommand{\Ae}{A_{\rm e}}

\newcommand{\Nh}{N_{\rm h}}

\newcommand{\rmstoa}{\sigma_{\rm TOA}}
\newcommand{\Save}{\langle S \rangle}

\title{Pulsars as Tools for Fundamental Physics \& Astrophysics}

\author{J.~M.~Cordes\address{Department of Astronomy and NAIC, 
			Cornell University, Ithaca, NY USA},
        M.~Kramer\address{University of Manchester, Jodrell Bank Observatory, 
                          Jodrell Bank, UK},
        T.~J.~W.~Lazio\address[NRL]{Naval Research Laboratory, 
                       Remote Sensing Division, Washington, DC USA}
                \thanks{Basic research in radio astronomy at the NRL
                is supported by the Office of Naval Research.},
        B.~W.~Stappers\address{ASTRON, Dwingeloo, The Netherlands},
	D.~C.~Backer\address{Department of Astronomy, UC Berkeley,
		Berkeley, CA USA},
	S.~Johnston\address{School of Physics, University of Sydney,
		Sydney, NSW 2006, Australia}
}

\begin{document}

\begin{abstract}
The sheer number of pulsars discovered by the SKA, in
combination with the exceptional timing precision it can provide,
will revolutionize the field of pulsar astrophysics. The
SKA will provide a complete census of pulsars in both the Galaxy and
in Galactic globular clusters that can be used to provide a detailed
map of the electron density and  magnetic fields, the dynamics of the
systems, and their evolutionary histories. This complete census 
will provide examples of nearly every possible outcome
of the evolution of massive stars, including the
discovery of very exotic systems such as
pulsar black-hole systems and sub-millisecond
pulsars, if they exist. These exotic systems will allow unique tests
of the strong field limit of relativistic gravity and the equation of
state at extreme densities.  Masses of pulsars and their binary companions 
--- planets, white dwarfs, other neutron stars, and black holes --- will
be determined to $\sim 1$\% for hundreds of objects.     
With the SKA we can discover and time 
highly-stable millisecond pulsars that comprise
a pulsar-timing array for the detection of low-frequency 
gravitational waves. 
The SKA will also provide partial censuses
of nearby galaxies through periodicity and single-pulse detections, yielding
important information on the intergalactic medium.
\end{abstract}

%\begin{keyword}
%Galaxy: structure, interstellar medium, magnetic field, Stars: evolution,
%Pulsars: neutron stars, equation of state, 
%Gravity: strong fields, tests.
%\end{keyword}

\maketitle

\section{Introduction}

Neutron stars (NSs) are accessible to observation as pulsars and thus
provide our only means for probing the most extreme states of
matter in the present-day Universe, which in turn
will enable a vast
range of transforming science goals to be addressed, among which are
\begin{itemize}
\itemsep -3pt
\item Strong-field tests of gravity including
the Cosmic Censorship Conjecture and the no-hair
theorem of BHs.
\item Detection of a cosmological gravitational wave background.
\item Mapping the complete structure of the Milky Way and 
revealing properties of the Galactic Center.
\item Probing the intergalactic medium in new ways.
\item Identifying the equation-of-state of super-dense matter. 
\item Quantifying the roles of magnetic fields and turbulence in
core-collapse physics.
\item Understanding the superfluid interiors and relativistic
magnetospheres of NS.
\item Unraveling the evolutionary and dynamical histories and properties
of all Galactic globular clusters.
\item Discoveries of extrasolar planets.
\end{itemize}

Internal densities $\sim$ ten times nuclear density have not
existed since the Universe was about 1 ms old.  The actual state of
matter in the core of a NS is presently not known.  It may
consist of de-confined quark matter or hyperonic matter produced in a
phase transition that occurred during or shortly after the core
collapse of the progenitor star.  Intermediate regions of the NS
consist of neutrons and trace protons in, respectively, superfluid and
super-conducting states achieved after the NS cooled to about
$10^9$ degrees.  The outer regions include an $\sim 1$ km thick crust
composed of iron-like nuclei surrounded by an ocean about 1 cm deep \cite{st83}.
The magnetic field anchored to the crust and extending to interstellar
space is sufficiently strong that it elongates the atoms comprising the
crust.  The surface gravity, about $10^9$ times that of the Earth's,
is the largest of any object in the Universe subject to observation,
and corresponds to a gravitational redshift $\sim 0.3$.

While NS gravity is strong, radio pulsars and probably most NSs
are even more extreme electromagnetically.  By virtue of spin
periods $P\sim 1$s and magnetic fields $B\sim 10^8-10^{14}$ Gauss, the
electric force on a surface proton $\sim 10^{11}$ times larger
than the gravitational force.  Voltage drops $\sim 10^{12}$ volts
across the magnetosphere accelerate particles that can radiate across
the entire electromagnetic spectrum. 
This makes some pulsars visible in every astronomical window
\cite{tho00b}. 
Most of what is to be learned from pulsars 
requires observations of their radio emission, often providing a
unique source of information, and otherwise  providing information
that complements multiwavelength
studies, particularly at high energies. 

The feature that gives pulsars their name --- pulsed radio emission ---
allows most NSs to be detected at levels well below the sky
confusion limit and also provides the means for using pulsars as
physics laboratories.  Coherent radio emission is
associated with the collimation of the flow of particles at the poles
of the large-scale magnetic field in combination with relativistic
beaming. The spin-driven sweep of the beam across the line of
sight then provides the distinctive pulsation of the electromagnetic
signal.  

The suitability for using pulsars as clocks depends on the regularity
by which the spin evolves with time. Spin noise in some objects, which
evidently reflects activity within the crust and superfluid, is large
enough to mask many of the physical effects of interest that provide
only subtle timing signatures. However, spin noise itself, including
rapid spinups (glitches), is itself valuable information on NS interiors.  
All pulsars are stable enough
so that timing measurements can yield fundamental parameters such as
the period vs. epoch, $P(t)$ and its time derivative $\Pdot$,  along with
the dispersion measure DM. Objects with the narrowest pulses, the
shortest periods, and the most stable rotation rates --- millisecond
pulsars --- yield the greatest opportunities for exploring 
relativistic gravity.  Extrinsic gravitational effects include
perturbations of pulse arrival times from the direct influence of
companion stars on space-time and from evolution of the orbit in the
non-Newtonian gravity.  The latter causes binary pulsars and their
orbits to precess whilst their orbit decays owing to loss of energy to
gravitational radiation.  Besides being sources of gravitational wave
emission \cite{pet64}, pulsars also lend themselves as detectors of long-period
gravitational waves that are cosmological in origin \cite{det79}.

The population of isolated and binary pulsars is of great interest
because their phase-space distribution and overall numbers reflect the
rate at which NSs are born in Type II supernova explosions, how the
explosions themselves produce the runaway velocities of NSs, and how
NSs are influenced by accretion in those rare binary systems that
survive the explosions. Unlocking the vast population of active
radio pulsars therefore allows us to study the star formation 
history and evolution of massive stars, aspects of binary evolution,
core collapse physics, as well as the movement of the high-velocity
population of pulsars in the Galactic gravitational potential.

With this enormous range of fundamental physics accessible through
pulsar observations, the pulsar field has been extremely
fruitful in the 37 years since their discovery, as evidenced by the
awarding of two Nobel prizes, one to the original discovery of pulsars
\cite{hbp+68}, the other to the discovery of the first NS-NS binary
pulsar \cite{ht75} that allowed inference of gravitational radiation
in accord with Einstein's General Theory of Relativity \cite{tfm79}.
Nonetheless, the field has a great deal more to contribute to our
knowledge of fundamental issues in physics and astrophysics.

We describe some of the applications of pulsars made possible by the
Square Kilometer Array project in the following Sections.  In
particular, we have identified {\it Strong-Field Tests of Gravity
Using Pulsars and Black Holes} as a key science area for the SKA that
is outlined in some detail in \S\ref{sec:psr.grtests} and in particular in
Kramer et al. (this volume).
%Chapter~\ref{chap:level0:pulsar}.

To enable this programme of research outlined below, the SKA must have
capabilities that allow its enormous collecting area to be used in a
variety of observing modes. Some of these imply usage as a huge,
effective single dish, while others exploit the array aspects of the
telescope.  As a mantra for pulsar research, the SKA must allow us the
following activities on pulsars: find them, time them and VLBI them.

%%%%%%%%%%%%%%%%%%%%%%%%%%%%%%%%%%%%%%%%%%%%%%%%%%%%%%%%%%
%%%%%%%%%%%%%%%%%%%%%%%%%%%%%%%%%%%%%%%%%%%%%%%%%%%%%%%%%%
\section{The Cosmic Census for Pulsars}
\label{sec:psr.cosmic_census}

Pulsars are of great utility, no matter where we find them. So far,
most known pulsars are Galactic, residing in or near the disk of the
Galaxy or in globular clusters.  A small number of pulsars is known in
the Magellanic clouds; however radio pulsars have not yet been detected in
more distant galaxies, although a few NS in accreting systems have
been seen in M31 via their X-ray emission \cite{kaa02}.  In the following we
summarize why it is important to discover more pulsars, both locally
and in other galaxies, as well as in particular regions of the Milky
Way.

\subsection{Galactic Census}
\label{sec:psr.galactic_census}

\begin{figure}
\includegraphics[width=7.3cm]{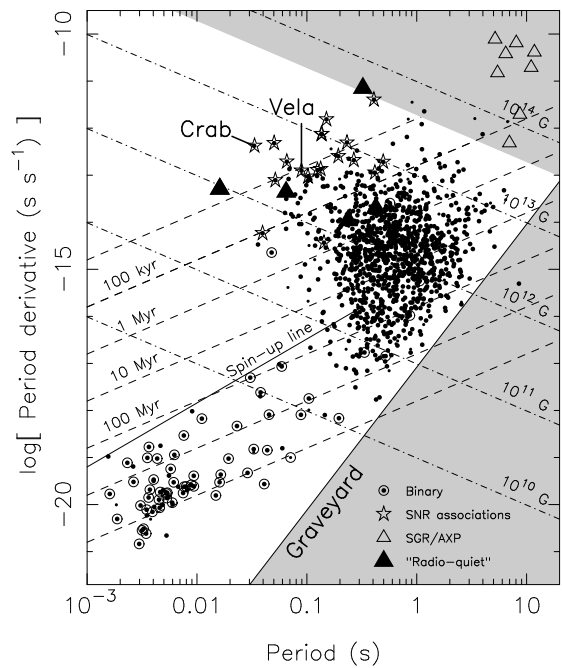}
\vspace{-0.7cm}
\caption{\footnotesize
\label{fig:psr.ppdot} The $P-\Pdot$ diagram for radio pulsars and magnetars.
This is a scatter plot of pulse period and period derivative for pulsars in the
current (Apr 2004) public ATNF catalogue.  The circled points are
pulsars in binary systems while objects in the upper right part of the
diagram have derived surface magnetic fields exceeding $\Bq =
4.4\times 10^{13}$ Gauss (see \S\ref{sec:psr.relphysics}).  The
``spin-up'' line denotes the terminal period for objects moving to the
left in the diagram owing to accretion that become ``recycled''
pulsars once accretion ceases.  The ``graveyard'' marks the locations
where radio pulsars emit less radiation or turn off entirely, possibly
associated with the cessation of electron-positron (e$^{\pm}$)
pair-production cascades or with saturation of the radio luminosity at
some fraction of the total spin-down energy loss rate.  }
\end{figure}

{\em Why perform a full-Galactic census?}  The Galactic Census is
essential in providing the laboratories for a wide range of pulsar
applications that will be discussed in more detailed in the following
sections.  The larger the number of pulsar detections, the more likely
it is to find rare objects that provide the greatest opportunities as
physics laboratories.  These include binary pulsars with black hole
companions (see \S\ref{sec:psr.grtests}); binary pulsars with
orbital periods of hours or less that can be used for fundamental tests of
relativistic gravity (see \S\ref{sec:psr.grtests}); MSPs that can be
used as detectors of cosmological gravitational waves (see
\S\ref{sec:psr.gw}); MSPs spinning faster than 1.5 ms,
possibly as fast as 0.5 ms, that probe the equation-of-state under
extreme conditions (see \S\ref{sec:psr.eos}); hypervelocity pulsars
with translational speeds in excess of $10^3$ km s$^{-1}$, which probe
both core-collapse physics and the gravitational potential of the
Milky Way (see \S\ref{sec:psr.supernovae}); and objects with unusual
spin properties, such as those showing discontinuities (``glitches'')
and apparent precessional motions (both isolated pulsars showing
`free'' precession and binary pulsars showing geodetic precession)
(see \S\ref{sec:psr.eos}).

The second reason to perform a full Galactic census is that the large
number of pulsars can be used to delineate the advanced stages of
stellar evolution that lead to supernovae and compact objects. In
particular, with as large a sample as possible, we can determine the
branching ratios for the formation of canonical pulsars and magnetars.
We can also estimate the effective birth rates for MSPs and for those
binary pulsars that are likely to coalesce on time scales short enough
to be of interest as sources of periodic, chirped
gravitational waves. The latter population can be directly compared
to the results of gravitational wave detectors, which will have been
operating for a sufficiently long time by the time the SKA is
switched on.

The third reason is that a maximal pulsar sample can be used to probe
and map the interstellar medium (ISM) in a nearly complete way.
Measurable propagation effects include dispersion, scattering, Faraday
rotation, and HI absorption that provide, respectively, line-of-sight
integrals of the free-electron density $n_e$, of the fluctuating
electron density, $\delta n_e$, of the product $B_{\parallel} n_e$,
where $B_{\parallel}$ is the LOS component of the interstellar
magnetic field, and of the neutral hydrogen density.  The resulting
measures are DM, SM, RM and $N_{\rm HI}$, 
\begin{eqnarray}
 \DM&=& \int_0^D ds\, \nel, \quad \SM= \int_0^D ds\, \cnsq, \\
 \RM&=& \int_0^D ds\, \nel \Bpar,\quad N_{\rm HI} = \int_0^D ds\, n_{\rm HI}, 
\end{eqnarray}
 involving quantities
that can also be studied by other means, for instance by HI
observations with the SKA.  The determination of these observables for
a large number of independent line-of-sights for pulsars 
will enable us to construct a complete map of the Galaxy 
(see \S\ref{sec:psr.ism}).

With a Milky Way birth rate that currently may be as large as
$10^{-2}$ yr$^{-1}$, about $10^8$ NSs have been formed over the
lifetime of the Galaxy, and probably many more because the
star-formation was most likely higher in the past. Most NSs are inert,
their radio emission having shut off long ago, and up to about half of
them will have left the Galaxy owing to their large space velocities.
Of the known pulsars (see Fig.~\ref{fig:psr.ppdot}), 
we can identify several subclasses:
\begin{enumerate}
\item {\it Canonical pulsars:} These pulsars, like those first
discovered, have present-day spin periods ranging from tens of
milliseconds to 8 s and surface magnetic field strengths $B \sim
10^{12\pm 1}$ G. They are often thought to be born with periods $\sim
10$ ms, though evidence suggests that some objects are born with
periods longer than 0.1 s \cite{klh+03}.  In the standard picture 
of NS formation,
{\it all} pulsars start as canonical pulsars. 
In the $P-\Pdot$ diagram of Fig.~\ref{fig:psr.ppdot} most of these pulsars
are located at $P\sim 1$s and $\Pdot\sim 10^{-15}$.
{\it Young pulsars} are especially important members of this class because
they are associated with supernova remnants and often show copious
numbers of glitches.
\item {\it Modestly recycled pulsars:} are objects in binaries that survived
a first SN explosion and subsequently accreted matter that spun-up the
pulsar and reduced the effective dipolar component of the magnetic field.
Accretion is terminated in these objects by a second supernova explosion
that may or may not disrupt the binary.  Those that survive are seen today
as relativistic NS-NS binaries.  
Evolutionarily, it is possible that 
some surviving binaries include
black-hole companions.
In the $P-\Pdot$ diagram of Fig.~\ref{fig:psr.ppdot} these pulsars
are typically located around $P\sim 30$ ms and $\Pdot\sim 10^{-18}$.
\item {\it Millisecond pulsars (MSPs):} 
objects in binaries that survive the first
SN explosion and in which the companion object eventually
evolves into a white dwarf.
The long, preceding
 accretion phase spins the pulsar up to millisecond periods while
attenuating the (apparent) dipolar field component to $10^8 - 10^9$ G.
The consequent small spin-down rates seem to underly the high timing
precision of these objects and imply spin-down time scales that
exceed a Hubble time in some cases.
In the $P-\Pdot$ diagram of Fig.~\ref{fig:psr.ppdot} these pulsars
are typically located around $P\sim 5$ ms and $\Pdot\sim 10^{-20}$.
Evolutionary scenarios that produce recycled pulsars and MSPs are discussed 
in \cite{bvdH91}.
\item {\it Strong-magnetic-field pulsars:}  Recently discovered radio
pulsars have inferred fields $\gtrsim 10^{14}$ G
\cite{ckl+00,msk+03}, rivalling those 
inferred for ``magnetar'' objects identified through their X-and-$\gamma$
radiation that seems to derive from non-rotational sources of energy. 
The relationship between magnetars and these high-field radio emitting
pulsars, whose radiation derives solely from spin energy, is not yet known.
In the $P-\Pdot$ diagram of Fig.~\ref{fig:psr.ppdot} these radio pulsars
are typically located around $P\sim 5$s and $\Pdot\sim 10^{-13}$.
\end{enumerate}

\begin{figure}
\includegraphics[width=7.3cm]{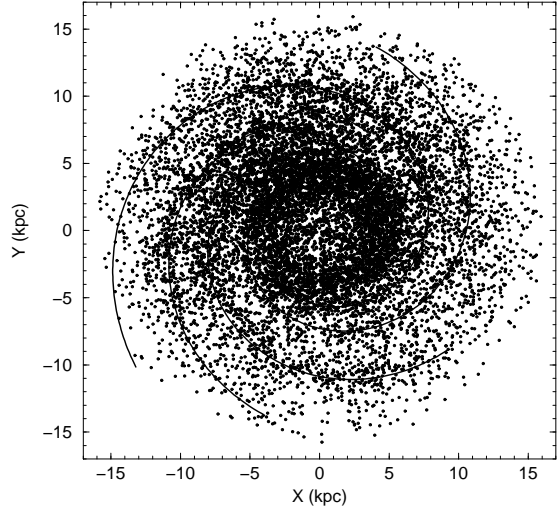}
\vspace{-0.7cm}
\caption{\footnotesize
\label{fig:psr.sky} Simulated Galactic pulsar population discovered
in a SKA survey of the enitre sky. The $\sim 20,000$ pulsars
shown together with the spiral arms structure. 
The Galactic Center is located at the origin while the Sun
is at (0.0,8.5) kpc.}
\end{figure}

We envision a full-Galactic census of radio pulsars that aims to
detect at least half of the active radio pulsars that are beamed
toward Earth (Figure~\ref{fig:psr.sky}).  The typical lifetime of canonical pulsars $\sim 10$
Myr, where we define lifetime as the duration of the radio-emitting
phase that, for objects with $B\sim 10^{12}$ G, is the time needed for
a rapidly rotating object to reach the ``death band'' on the
right-hand side of the $P-\Pdot$ diagram.  At long periods short-ward
of the death band, where pulsars spend most of their detectable
lifetimes, the beaming fraction $\sim 20$\% (e.g. \cite{ec89}), 
so the fiducial birth
rate implies $\sim 2\times 10^4$ detectable pulsars in the Galaxy.
Non-canonical classes of pulsars add to these numbers only negligibly
because their effective birth rates are smaller by a factor
$10^{-4}$--$10^{-3}$.

%%%%%%%%%%%%%%%%%%%%%%%%%%%%%%%%%%%%%%%%%%%%%%%%%%%%%%%%%%

\subsection{Galactic Center}
\label{sec:psr.center_census}

The Galactic Center (GC) is an especially tantalising but exceedingly
difficult region to search for pulsars.  In many respects, the GC is
similar to an especially large globular cluster with regard to the
density of stars (cf.~\S\ref{sec:psr.cluster_census}). It differs
in that molecular material and, hence, star formation are both much
more prominent in the GC than in globulars.  Additionally, the $\sim
3\times 10^6\Msun$ black hole \cite{sog+02} that underlies the compact radio source,
Sgr A*, perturbs space time significantly and is fed episodically by
inspiraling gas and stars.  

{\em Why find pulsars in the GC?}  Any pulsars detected in or beyond
the GC that are viewed through the region are potentially unique
probes of the gas, magnetic field, and space-time of the GC region.
First of all, pulsars in the GC can be used to probe the magnetoionic
medium along the line-of-sight, including possible detection of the
inner scale for electron density turbulence (see
\S\ref{sec:psr.mapcenter}).  Second, the initial mass function and overall
stellar evolution in the GC is likely to be very different from the
rest of the Galaxy.  The number and ages of pulsars and their binary
membership will provide clues about these areas (see
\S\ref{sec:psr.mapcenter} and \S\ref{sec:psr.population}). Finally, the large
stellar density offers the possibility of finding pulsars with stellar
black hole companions, allowing unprecedented tests of gravitational
theories in the strong field limit and the study of black hole
properties.  Similar studies will be possible for Sgr A*, the
supermassive black hole in the center, if pulsars in compact orbits
around the black hole are found (see \S\ref{sec:psr.grtests}).  Pulse
timing measurements may provide the possibility of measuring the
spin of the GC's central black hole.

At present, {\it none} of the known pulsars is within or beyond 
the GC.  The primary reason is
that an especially intense scattering region lies between us and the
GC at close proximity ($\sim 100$ pc) to the GC \cite{lc98b}.  
The large scattering has been known
since shortly after Sgr A* itself was discovered 30 years ago
and it has been probed through angular diameter measurements of 
OH/IR masers in the GC region and through surveys for background AGNs.
The implication for pulsars is that, at a standard search frequency
of 1.4 GHz, a pulse emanating from the location of Sgr A* would be
broadened to $\sim 300$ s owing to multi-path 
propagation.\footnote{
Pulsars located closer to us than Sgr A* but still behind the screen 
are  scattered less than this
while pulsars beyond the GC viewed through the screen are more scattered.}
  
\begin{figure}
\includegraphics[width=7.5cm]{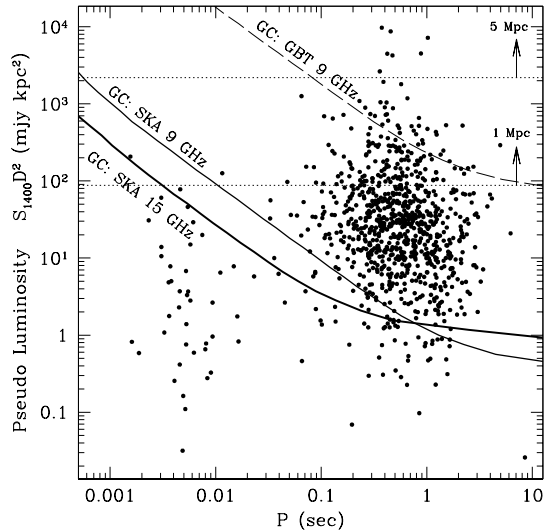}
\caption{\footnotesize
\label{fig:psr.ngcdetect}
Detectability of pulsars in the Galactic Center and in other galaxies
is shown in this plot of 
``pseudo luminosity'' $\Lp = S_{1400} D^2$ vs. spin period for pulsars
in the current (Apr 2004) public ATNF catalogue;  $S_{1400}$ is the
period-averaged flux density at 1.4 GHz.  Given duty cycles for
pulsars ranging from 0.01 to 0.7, the peak flux density can be a large
multiple of the period-averaged flux density.  Solid and dashed lines
mark the minimum detectable spin period $P_{\rm min}$ for pulsars at
the Galactic Center that are detectable with the GBT at 9 GHz and the
SKA at 9 and 15 GHz. Scattering is assumed to be in a filled region
centered on Sgr A* with $1/e$ cylindrical radius of 0.15 kpc, as
included in the NE2001 model. For pulsars at the location of Sgr A*,
the dispersion and scattering measures are
$\DM_{\rm GC} = 1577$ pc cm$^{-3}$ and
$\SM_{\rm GC} = 10^7$ kpc m$^{-20/3}$.
The curves are based 
on harmonic sums of the Fourier power spectra of dedispersed
time series. They correspond to surveys at these frequencies but
have all been scaled to the equivalent sensitivity at 1.4 GHz assuming
that the pulsar flux densities scale as $\propto \nu^{-2}$.
 A bandwidth of 800 MHz was assumed in all cases.
Integration times of 1.5 hr and $10^4$ s were used for the GBT and the
SKA, respectively. In addition, the SKA curves assume that the full
sensitivity of the SKA is available.  Dotted horizontal lines show the
minimum values for $\Lp$ needed for detection at 1.5 GHz in
periodicity searches of pulsars at distances of 1 Mpc and 5 Mpc.  The
dotted lines assume detectability is independent of spin period, which
will not be the case if there is significant scattering along the line
of sight. 
} 
\end{figure}

The pulse broadening time $\tau_d$ scales with the {\it observed}
angular diameter $\theta_d$ as $\tau_d =
{W_{\tau}D\theta_d^2}/{2c}$, where $D$ is the source distance and
$W_{\tau}$ is a geometrical weighting factor that takes into account
the location of the scattering region relative to the source and
observer.  For the GC, the weighting factor is very large because the
region is much nearer the GC than the observer.  For scattering by
fluctuations in the electron density, $\theta_d \propto \nu^{-2}$ and
thus the broadening time is a very strong function of frequency,
$\tau_d \propto \nu^{-4}$.

To combat pulse broadening, observations at higher frequencies are
therefore needed that exploit the strong $\nu^{-4}$ dependence.  For
example, 10 GHz observations yield $\sim 0.11$s broadening, small
enough to allow detection of longer period pulsars.  However, pulsar
spectra are power-law in form, often with steep dependences $\propto
\nu^{-\alpha}$ with $\alpha$ ranging from 0 to 3 and $\langle \alpha
\rangle \approx 1.5$.  While ongoing searches may yield detections of
a few pulsars in the GC, some of which may provide important probes of
the region, the sensitivity of the SKA is clearly needed 
at high frequencies
to detect a meaningful sample of pulsars in the GC,  as demonstrated
in Figure~\ref{fig:psr.ngcdetect}.
The figure shows the radio luminosity (defined as $\Lp = $ flux
density $\times$ distance$^2$) at 1.4 GHz plotted against spin period
along with detection threshold lines for the Green Bank Telescope (GBT)
and for the SKA.

The maximum detectable distance is
$$
\Dmax = \left( \frac{\Lp}{\Sminone} \right)^{1/2} \Nh^{1/4}, 
$$
where $\Sminone = m\Ssys / \sqrt{2BT}$ is the single-harmonic
threshold with  $m$ = the threshold
signal-to-noise ratio, and $\Nh$ is the number of harmonics detected
in the search power spectrum.  The minimum luminosity for $m=10$
is thus
$$
{\Lp}_{\rm min} = 87.5 \, {\rm mJy\, kpc^2}
	\left( \frac{\Dmax}{1\, {\rm Mpc}} \right)^2
	\left( \frac{\Nh}{16} \right)^{-1/2}.
$$
For short period pulsars, the number of harmonics detected will
be less than the 16 number assumed in this equation.

Also shown in Figure~\ref{fig:psr.ngcdetect} 
are detection curves for the minimum detectable period $P_{\rm
min}$ vs. $\Lp(1.4 \,{\rm GHz})$.  These are shown for pulsars at the
Galactic Center that are detectable with the Green Bank Telescope
(GBT) at 9 GHz and with the SKA at 9 GHz and 15 GHz.  These curves
were calculated for pulsars near the location of Sgr A* and take into
account pulse broadening from scattering and the spectral dependence
of the pulsar flux density using a conservative value for spectral
index, $\alpha = 2$.  The detectability was thus calculated for
surveys at these frequencies but has been scaled to the equivalent
sensitivity at 1.4 GHz.  The calculations assume that the full
sensitivity of the SKA is available.  If only a fraction $f_c$ can be
used in a blind survey using a ``core'' array, then the curves will
move upward by an amount $\log f_c$.

%%%%%%%%%%%%%%%%%%%%%%%%%%%%%%%%%%%%%%%%%%%%%%%%%%%%%%%%%%

\subsection{Globular Clusters}
\label{sec:psr.cluster_census}

Globular Clusters hold vast reservoirs of MSPs, which are formed
at a rate per unit mass which is at least an order of magnitude higher
than in the Galactic disk. The reason for this overabundance, which
may be even higher than presently seen due to the escape of high velocity
objects, is thought to be the formation of binaries via two-
and three-body encounters in the high density environments of Globular
Clusters.

{\em Why perform a full census of globular clusters?}  The greatly
increased likelihood that many of these MSPs in globular clusters will
have undergone some form of dynamical interaction means that the
chances of finding exotic binaries, such as the long sought-after
MSP-black hole system, are perhaps highest in globular clusters (see
\S\ref{sec:psr.grtests}). Furthermore the pulsars in each globular
cluster will provide us with exceptional probes of the history and
evolution as well as its present properties, including the dynamics,
gas content, accurate distances and proper motion (see
\S\ref{sec:psr.mapcluster}).  Moreover, an open question is whether
globular clusters contain massive black holes in their centers
\cite{mpzh04,gfr04}, or
possibly even binary black holes \cite{cmp03}. 
Pulsar timing with the precision
achievable with the SKA will provide us with a tool to reveal whether
such systems are present. This can be achieved either by probing the
inner most regions of the cluster to reveal velocities or
accelerations expected due to the presence of a black hole, or by
using the pulsars in the cluster as an in situ gravitational wave
detector sensitive to the presence of binary black holes
(see \S\ref{sec:psr.gw}).

At present there are 76 pulsars known in 23 globular
clusters. Not all clusters are equally rich, and it is not clear what
determines the number of active pulsars in a given cluster.  Based on
our knowledge of the luminosity function of the MSPs in the Galactic
disk combined with the continuum radio emission and the numbers of
MSP-like X-ray sources associated with globular clusters, the total
population is likely to be at least two orders of magnitude more than
this. 

The remaining pulsars in these globular clusters are difficult
to find with present instruments due to a combination of the intrinsic
low luminosity of MSPs and the typically large distances to globular
clusters. Furthermore, most pulsars are in compact binary orbits.
Given the presently long integration times needed
to detect the pulsars, the pulse signal is often smeared out due to
the Doppler effect. Recent developments of sophisticated techniques to
correct for this smearing have resulted in an increase in the
number of systems known, but blind searches are often too
computationally expensive and the vast majority of systems are still
too weak for detection.

Targeted SKA surveys of all Galactic globular clusters will enable us
to detect all of the pulsars contained therein which are beamed in our
direction. This complete census will uncover large numbers of pulsars
with which to study their formation, evolution, spin parameters,
binary nature and emission properties. 

%%%%%%%%%%%%%%%%%%%%%%%%%%%%%%%%%%%%%%%%%%%%%%%%%%%%%%%%%%

\subsection{External Galaxies}
\label{sec:psr.extragal_census}

Pulsars beyond the disk of the Milky Way are known only in globular
clusters and in the Magellanic Clouds owing to their intrinsic
faintness.  With the SKA, galaxies in the local group, including M31
and M33, are within reach using periodicity searches while giant
pulses like those seen from the Crab pulsar can be detected from
galaxies out to at least 5 Mpc.

{\em What is the importance of detecting pulsars in other Galaxies?}
Pulsars likely to be detected will be young pulsars with high
luminosities that can be correlated with catalogs of supernova
remnants and will yield estimates of the star-formation rate and the
branching ratio for supernovae to form spin-driven pulsars as opposed
to magnetars and blackholes.  Extragalactic pulsars will also provide
information about the magnetoionic media along the line of sight
through determination of \DM, \SM, and \RM.  Unambiguous study of the
intergalactic medium in the local group requires removal of
contributions to these measures from the foreground gas in the Galaxy
and gas in the host galaxy.  The more pulsars detected in a galaxy,
the more robust this removal will be.

Extragalactic pulsars can be found through blind surveys for both
periodic sources and individual giant pulses.
Additional successes will follow from targeted surveys of individual
supernova remnants in the nearest galaxies.  In \S\ref{sec:psr.searches} we
discuss the requirements on sensitivity and field of view (FOV).

Figure~\ref{fig:psr.ngcdetect} includes detection lines for surveys at 1.4
GHz for pulsars at 1 Mpc and 5 Mpc assuming that the full SKA
sensitivity is available.  Full sensitivity would apply to targeted
searches of, for instance, supernova remnants in nearby galaxies.
However, $\sim$full-FOV blind surveys will provide only a fraction
$f_c$ corresponding to a core array.  With full sensitivity, a
reasonable fraction of the luminosity function can be sampled out to 1
Mpc and a few pulsars can be detected to 5 Mpc.

\newcommand{\SCN}{S_{\rm CN}}
\newcommand{\DCN}{D_{\rm CN}}
\newcommand{\Ssyso}{S_{\rm sys_0}}

Giant pulses from the Crab pulsar serve as a useful prototype for
estimating detection of strong pulses from nearby galaxies.  The
strongest pulse observed at 0.43 GHz in one hour has $S/N_{\rm max}
= 10^4$ -- even with the system noise dominated by the Crab
Nebula.  For objects in other galaxies, the system noise is
dominated by non-nebular contributions, implying that
the S/N in this case would have increased by a factor of about 300.
We can estimate the maximum distance of
detection at a specified signal-to-noise ratio, $(S/N)_{\rm det}$ as
\be
D_{\rm max}
        &\approx&  \frac{ 1.6\,{\rm Mpc} }
        { \sqrt{(S/N)_{\rm det}/5} }
	\left (\frac{f_c A_{\rm SKA}}{A_{\rm Arecibo}} \right)^{1/2},
\ee
where $f_c A_{\rm SKA}$ is the SKA's collecting area
that can be used for a giant pulse survey and $A_{\rm Arecibo}$ is
Arecibo's effective area at 0.43 GHz.  The usable area for the SKA
in a blind survey will be limited by what fraction $f_c$ of
the antennas are directly connected to a central correlator/beamformer.
For $f_c = 1$, $A_{\rm SKA} / A_{\rm Arecibo} \approx 20$, the standard
one-hour pulse seen at Arecibo could be detected out to 7.3 Mpc.

%%%%%%%%%%%%%%%%%%%%%%%%%%%%%%%%%%%%%%%%%%%%%%%%%%%%%%%%%%

\section{Fundamental Physics \& Astrophysics}
\label{sec:psr.applications}

Having discovered a large sample of pulsars in the census of the
Milky Way, the Galactic Centre, Globular Cluster and external
galaxies, a vast range of fundamental problems in physics
and astrophysics can be studied. We highlight some of those
in the following.

\subsection{Tests of Theories of Gravity}
\label{sec:psr.grtests}

The theory of General Relativity (GR) has to date passed all
observational tests with flying colours. Nevertheless, one of the most
fundamental questions remaining is {\em whether Einstein's theory is
the last word in our understanding of gravity or not}.  Solar system
tests of GR are made under weak-field conditions, illuminated by the
small orbital velocities and gravitational self-energies
involved (e.g.~the self-energies for the Sun and the Earth, expressed
in units of their rest-mass energy, are $\epsilon_{\rm Sun}\sim
-10^{-6}$ and $\epsilon_{\rm Earth}\sim -10^{-10}$, whilst for a
pulsar and black hole (BH) we find $\epsilon_{\rm PSR}\sim -0.2$ and
$\epsilon_{\rm BH}=-0.5$). Therefore, none of these tests of
gravitational theories can be considered to be complete without
probing the strong-field limit, which is done best and with amazing
precision using binary radio pulsars \cite{de98}. However, even the
existing binary pulsar tests only begin to approach the strong-field
regime and the discovery of more extreme binary systems is
required. Indeed, the SKA is the only instrument which promises to
probe the strong-field limit via the discovery and timing of such
systems, in particular that of a pulsar orbiting a BH. The promises
provided by the discovery of such particular systems are discussed in
detail in a companion key science article (Kramer et al., this
volume), where we detail how pulsars can
be used to test the ``Cosmic Censorship Conjecture'' and the
``No-hair'' theorem for the description of BHs. Here, we concentrate
on general aspects of tests of gravitational theories using pulsars
timed with the SKA.

Since NSs are very compact massive objects, double neutron star (DNS)
binaries can be considered as ideal point sources.  Finding and timing
DNSs in tight binary orbits -- ideally close to coalescence and
thus  emitting
strong gravitational waves -- provide stringent tests of theories of
gravity in the strong-gravitational-field limit.  Tests can be
performed when a number of relativistic corrections to the Keplerian
description of the orbit, the so-called ``post-Keplerian'' (PK)
parameters, can be measured. For point masses with negligible spin
contributions, the PK parameters in each theory should only be
functions of the a priori unknown NS masses and the well measurable
Keplerian parameters. With the two masses as the only free parameters,
the measurement of three or more PK parameters over-constrains the
system, and thereby provides a test ground for theories of gravity
\cite{dt92}.
In a theory that describes the binary system correctly, the PK
parameters produce theory-dependent lines in a mass-mass diagram,
which compares the masses of the two neutron stars, that all intersect
in a single point.

The best example for such tests is currently given by the double
pulsar system PSR J0737$-$3039 where five PK parameters are available
for tests, in addition to the theory independent mass ratio of the two
NSs \cite{lbk+04}.  For the relativistic binary systems that will be
discovered and timed with the SKA, one would be able to routinely
measure five or more PK parameters, severely over-constraining the
system. In particular, more double pulsar systems will be found, and
PK parameters that are currently impossible (or extremely difficult)
to measure, such as those arising from aberration effects and geodetic
precession, would become accessible with the SKA, providing also full
3-D information about the orientation of these binary systems.

Most importantly, current PK parameters are only measured to the lowest
post-Newtonian approximation. The timing precision achievable with the
SKA means that higher-order corrections are likely to become
important, demanding the development of a timing formula that is
accurate to at least second post-Newtonian order. For instance,
corrections would need to account for the assumptions currently made
in the computation of the Shapiro delay (see \S\ref{sec:psr.eos}) that
gravitational potentials are static and weak everywhere
\cite{wex95a,ks99,kop03}.  In addition, other effects such as those related
to light bending and its consequences will become important where an
additional signal would be superposed on the Shapiro delay as a
typically much weaker signal, which arises due to a modulation of the
pulsars' rotational phase by the effect of gravitational deflection of
the light in the field of the pulsar's companion \cite{dk95}.

Like the DNS binaries, most NS-white dwarf (NS-WD) binaries can also
be considered as pairs of point masses.  Significant measurements of
PK parameters can only be obtained in a few cases so far, since
relativistic effects for NS-WDs are generally much smaller than for
DNS binaries (e.g.~\cite{vbb+01,bokh03}).  However, with the SKA, PK
parameters can be determined for many more systems.  Such measurements
can be similarly valuable since NS-WD systems test different aspects
of gravitational theories.  For instance, as shown recently
\cite{esp04}, the tests provided by the PSR-WD system J1141$-$6545 are
more constraining for the class of tensor-scalar theories than those
tests provided by the double pulsar. The reason is that unlike GR,
some alternative theories of gravity, such as
tensor-scalar theories, predict effects that depend strongly on the
difference between the gravitational self-energy of two orbiting
bodies, which is large in NS-WD systems. Using such systems, one is
able to probe possible violation of conservation of momentum,
equivalence principles and expected Lorentz- and positional
invariances \cite{sta03}.  A manifestation of such effects would, for
instance, be measurable in the detection of gravitational {\em dipole}
radiation where as GR only predicts {\em quadrupole} emission.

The computer power available when the SKA comes online will enable us
to do much more sophisticated searches in parameter space,
including full searches in acceleration due to binary motion, than
possible today, and the SKA sensitivity allows much shorter
integration times, so that searches for compact binary pulsars will no
longer be limited. Hence, the combination of SKA sensitivity and
computing power means that the discovery rate for relativistic
binaries is certain to increase beyond the number of at least a
hundred compact binary systems that we can expect from an
extrapolation of the present numbers.

In summary, tests of gravity affordable with the SKA will not simply be a
continuation of the present tests at higher precision levels, but the
better sensitivity and timing precision will enable us to perform
genuinely new tests. These prospects arise from the assumption that
the timing precision can be improved by two or even three orders of
magnitude over the current standards. However, as we discuss in some
detail in \S\ref{sec:psr.timing}, this requires not only an increase in
raw sensitivity but also the ability to correct for propagation effects
from intervening plasmas and  
sufficient polarization purity and (quasi-)
simultaneous multi-frequency observations. Moreover, the intrinsic
phase-jitter of pulsars will prevent the timing precision 
of some MSPs from reaching the
theoretical limit given by radiometer noise, so that the application
of correction schemes needs to be considered on a case-by-case
basis. However, the current timing precision of 
a number of MSPs appears to be 
limited by telescope sensitivity, as opposed to systematic effects
(e.g.~\cite{van03}), suggesting that 
much improved timing precision can be achieved for a few objects
and that new, unique tests
of relativistic gravity will be enabled through greater telescope sensitivity.

It is important to note that many of these experiments require
corrections of the measured parameter values for kinematic
effects. For instance, the precision of GR tests achievable with PSR
B1913+16 is limited by the accuracy to which the pulsar's motion and
acceleration in the gravitational potential of the Galaxy is known
\cite{wt03}. Observed values for parameters like the
orbital decay rate, $\dot{P}_{\rm b}/P_{\rm b}$, 
or changes in the semi-major
axis, $\dot{x}/x$, are affected by a kinematic Doppler term given by 
\cite{dt92}
\begin{equation}
    -\frac{\dot D}{D} =
           \frac{1}{c} \: \vec{K}_{\rm 0} \cdot (\vec{a}_{\rm PSR} -
  \vec{a}_{\rm SSB})
          + \frac{V_T^2}{c\: d}\; ,
\end{equation}
where $\vec{K}_{\rm 0}$ is a vector from the Solar System Barycentre
(SSB) towards the binary pulsar, $\vec{a}_{\rm PSR}$ and $\vec{a}_{\rm
SSB}$ are the Galactic accelerations at the location of the binary
system and the SSB. The last term including the transverse velocity,
$V_{T}$, and distance, $d$, to the pulsar is known as secular
acceleration or ``Shklovskii term'' \cite{shk70}. Correcting for this
term to derive the intrinsic values obviously requires precise
astrometric information like proper motion and distances which can be
derived using the VLBI capabilities of the SKA. In order to get precise
distances for pulsars across the Galaxy a final positional accuracy of
0.1 mas is needed at 5 GHz.

%%%%%%%%%%%%%%%%%%%%%%%%%%%%%%%%%%%%%%%%%%%%%%%%%%%%%%%%%%

\subsection{Structure of Neutron Stars \& Equation-of-State}
\label{sec:psr.eos}

The internal structure of NSs is complex. This question was
already addressed by Oppenheimer \& Volkov (1939)\nocite{ov39} long
before the discovery of pulsars. The structure depends sensitively on the
{\em equation of state}, i.e.~the relationship between density and
pressure.  Knowledge of the exact equation of state would allow us to
deduce most of the NS's physical properties, most notably
its radius for a given mass. With the SKA we can address this 
question from various angles.

Firstly, the study of the equation of state will benefit from the
vastly improved statistics of observed pulse periods. 
Periods of known pulsars  range from 1.5 ms to 8.5 s. The discovery of even
smaller rotational periods will provide significant insight into the
properties and stiffness of the ultra-dense liquid interior of NSs.
Even the case of a ``null result'', that no period shorter
than 1.5 ms will be found, will require a theoretical explanation, as
it requires the existence of a limiting period not far away from the
currently observed value but larger than our current theoretical
limits. At the other extreme end, the discovery of very long period
pulsars, isolated or in binary systems, will establish the connection
of radio pulsars to magnetars, Soft-Gamma ray Repeaters (SGRs) and
Anomalous X-ray Pulsars (AXPs), clarifying as to whether these are
distinct classes of NSs or simply different evolutionary
stages or end products (see also \S\ref{sec:psr.relphysics}).

\begin{figure}[h]
\includegraphics[width=7.5cm,angle=-90]{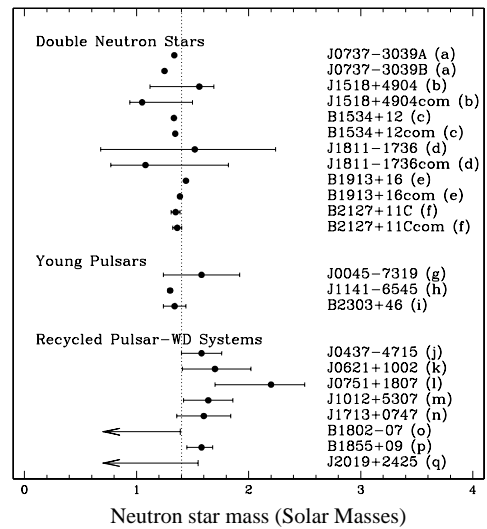}
\vspace{-0.7cm}
\caption{\footnotesize
\label{fig:psr.masses} Mass measurements of neutron stars obtained from
the timing of binary pulsars with corresponding uncertainties shown.
The canonical value of 1.4$\Msun$ is indicated by the vertical line.
While mildly recycled and young pulsars are broadly consistent with
this value, millisecond pulsars born in the process of a 
prolonged mass transfer appear to have somewhat larger masses.
Figure compiled by Stairs (2004).}
\end{figure}

Secondly, observations of relativistic effects in binary pulsars
allows precise determinations of the masses of the pulsar and its
companion. This is possible when either two or more PK parameters can
be determined (see \S\ref{sec:psr.grtests}), or when a Shapiro delay can
be measured.  The Shapiro delay in the arrival time of a pulse arises
from the curvature of space-time caused by the presence of a companion
star. Measuring the Shapiro time delay is therefore the most
straightforward way to obtain a direct measurement of the companion
mass and, hence, of the mass of the active pulsar from knowledge of the
total system mass (e.g.~by measuring a relativistic precession of the
orbit). Currently, the masses of about 20 neutron stars can be
determined (Fig.~\ref{fig:psr.masses}).  With hundreds of NS mass
measurements available, the range of observed masses can be studied
and related to the spin periods, promising to reliably establish the
maximum possible mass of observable NSs.  Moreover, the observed
values can be related to the evolutionary history of the binary
systems, studying the amount of matter accreted during a spin-up
process as an X-ray binary. Given that some millisecond
pulsars are very old, 
one may also be able to put limits on the evolution of NS
masses with age of the universe, so that variations in the gravitational
constant can be detected or constrained
\cite{tho96b}.  It is also worth pointing out that, in addition to
precisely measuring the distribution of masses of NSs, the same
observational techniques can be used to obtain mass measurements of
other compact pulsar companions.  The current timing precision usually
prevents this, but with the SKA it would be possible to perform such
measurements also for WD companions.
\cite{sta04}

Thirdly, the moment of inertia of NSs can be
determined when high precision timing observations allow the detection
of relativistic spin-orbit coupling in DNSs. This effect would be
visible most clearly as a contribution to the observed periastron advance
in a secular
\cite{bo75b} and periodic fashion \cite{wex95}.  
The size of the contribution depends on the pulsars' moments of
inertia, which therefore can be determined if two other PK parameters
can be determined with sufficient accuracy so that the effect of
spin-orbit coupling can be isolated \cite{ds88}.  The SKA timing
precision will be crucial in pushing the measurement accuracy to the
limits required to see this effect of second Post-Newtonian
(i.e.~$(v/c)^4$) order (cf.~\S\ref{sec:psr.grtests}).

Finally, observations of rotational instabilities in pulsars, 
which include rapid spin-ups followed by slow relaxation
--- {\em glitches} ---  and stochastic variations generically called
{\em timing noise}, 
probe the interior of NSs in a unique way that can be
considered ``neutron star seismology''. Observations of glitches
and their recovery provide strong evidence for the existence of a
fluid component inside the solid outer crust of the NS,
providing the basic picture of NS structure provided in the
introduction. This picture is based on only a small number of glitches
observed in even fewer pulsars (e.g.~\cite{sl96}).  The reason is that
glitches -- mostly observed in young pulsars -- are rare, so that their
timely detection is important to follow the revealing relaxation
process. With the SKA a large number of pulsars would be monitored
simultaneously, so that glitches can be observed as they happen. The
multi-beam capabilities of the SKA are essential for this study.

The results obtained from the analysis of the detected glitches can be
compared to the observations of pulsars that are found to be freely
precessing, since the time-scales for precession are probably related
to the coupling-strength between the liquid interior and the solid
crust. Currently, convincing evidence for free precession is found for
only a few pulsars \cite{sls00}, but SKA timing observations of all
pulsars discovered in the cosmic census will produce further examples.

Timing noise constitutes stochastic variations in pulse phase residuals
(e.g. after removing spin-down polynomial components and orbital
effects) that show non-stationary statistics and is thus
often described as having a ``red'' power spectrum.   
It is radio-frequency
independent (after any DM and scattering variations 
from the ISM are subtracted)  and appears to be caused by torque variations
acting on the NS crust.  It is possible that some or all timing noise
is associated with free precession induced by crust quakes that cause
misalignment of the angular momentum vector and the spin axis.  This
conjecture can be tested only by identifying more objects that show
pulse shape variations caused by wobble of the radio beam that accompanies
pulse phase variations.

%%%%%%%%%%%%%%%%%%%%%%%%%%%%%%%%%%%%%%%%%%%%%%%%%%%%%%%%%%
%%%%%%%%%%%%%%%%%%%%%%%%%%%%%%%%%%%%%%%%%%%%%%%%%%%%%%%%%%

\subsection{Pulsars as sources and detectors of gravitational waves}
\label{sec:psr.gw}

Pulsars are not only detectable across the whole electromagnetic
spectrum, but rotating and coalescing NSs are also among the
sources expected to be detected first with gravitational wave (GW)
detectors. 

The recent discovery of the J0737$-$3039 system suggests that Advanced
LIGO, operating on timescales similar to those for the SKA, will detect a few
coalescing binary pulsars each day \cite{bdp+03,kkl+04}. 
Moreover, it is anticipated that
J0737$-$3039 can act as a calibrating signal for the space
interferometer LISA. The SKA will discover many more such compact
relativistic binaries and will hence provide key information for a
location and study of these objects in a non-photonic window.

Moreover, deviations from a purely spherical shape will cause spinning
NSs to emit continuous GW emission that may even be detected with
current generations of GW telescopes like LIGO, GEO600 or VIRGO. The
faintness of the signal requires knowledge of the precise positions
and spin-frequencies of the pulsars.  With the majority of 
Galactic pulsars potentially being discovered with the SKA 
(see \S\ref{sec:psr.galactic_census}), sensitive targeted GW searches
are possible for a very large number of pulsars, so that the
actual shape of a NS can be probed.

As a result of the cosmic census, the SKA will also produce a dense
array of millisecond pulsars across the sky. Being timed to very high
precision ($\lesssim 100$ ns), these act as multiple arms of a cosmic
GW detector when a passing gravitational wave perturbs the
metric and hence affects the pulse travel time and the measured
arrival time at Earth
\cite{det79,fb90,rr95a}.  This ``device'' with the SKA at its heart
-- the so-called ``Pulsar Timing Array'' (PTA) -- will be sensitive to
GWs to frequencies of 1/observing time, hence $\sim$nHz.  The PTA
thereby complements the much higher frequencies accessible to Advanced
LIGO ($\sim$100Hz) and LISA ($\sim$mHz), and the extremely low
frequencies probed by polarization studies of the Cosmic Microwave
Background ($\sim 10^{-18}$ Hz). 

The measurement precision and accuracy of the pulsar clock is not
sufficient to detect the gravitational radiation of stellar-mass
binaries by the means of a PTA. However, super-massive black hole
binaries in nearby galaxies with orbital
periods of a few years would produce a periodicity in pulsar arrival
times with an amplitude of the order of 10 ns to 1 $\mu$s and in a
frequency range that is detectable with SKA timing
\cite{rr95a,lb01}. The light-travel time delay between the Earth and
the timed pulsars can, depending on the geometry, enable us to observe
a given massive binary system at two different epochs simultaneously.
A slow decay of the binary orbit hence results in both low and high
frequency components of the timing residuals, whereas the difference
in the frequencies of these components will depend on the orbital
decay rate.  As pulsar timing is more sensitive to lower frequencies,
the highest amplitude oscillations in the timing residuals will be due
to the delayed component.  This effect corresponds to the three-pulse
response occurring in spacecraft Doppler tracking experiments and the
multi-pulse response from time-delay interferometry used with LISA
\cite{jllw03}.

The short lifetime of such massive binaries reduces the chances
of detecting such systems. In contrast, the SKA can detect the signal
of a stochastic background of gravitational wave emission produced by
a large number of unresolved independent and uncorrelated events.
Another contribution to the GW background, but with a different
spectrum, is expected in some cosmological string theories. The SKA is
crucial in answering the question about the existence, nature and
composition of a stochastic GW background. The construction of a PTA to
detect such signals is one of the key-science projects of the SKA and
is presented in more detail in the contribution by 
Kramer, this volume.

\subsection{Relativistic Plasma Physics}

\label{sec:psr.relphysics}

The plasma physics occurring under
extreme conditions of very high densities, super-strong
magnetic fields and in the rather complex environment known as the
{\em pulsar magnetosphere} are only poorly understood.  This situation
remains after almost 40 years of pulsar research and is in spite of
the large number of studies and observations devoted to the
identification of the relevant physical processes 
(e.g.~\cite{mel00a}).  Solving this
``pulsar problem'' however offers deep insight into relativistic
plasma physics, and the SKA may provide us finally with the
observational facts to understand pair production processes, the
creation of radio and high-energy emission and the overall structure
of the pulsar magnetosphere.

{\em Radio Fluctuations:}
The size of the region of electromagnetic activity around active pulsars,
the magnetosphere, is essentially the light-cylinder radius,
\be
R_{\rm LC} = \frac{c}{\Omega} = \frac{cP}{2\pi} 
\simeq 4.77 \times 10^{4} \;    \mbox{km} \;
\left(  \frac{P}{\rm s} \right).
\ee
Radio emission altitudes are much smaller than $R_{\rm LC}$ and radio
emission is expected to originate as close to the NS surface as 10-100
stellar radii, produced by a relativistic pair plasma that is created
in the polar gap region \cite{kxj+97}.

While pulsars are renowned for their stable average pulse profiles
formed by the summing together of many pulses, single pulses provide
us with the rich detail that reflects the physics of the generation
and propagation of radio emission. The initially chaotic appearance of
single pulses can often be resolved to reveal a rich phenomenology
that includes: drifting sub-pulses, nulling, microstructure and giant
pulses which correspond to the building blocks of the observed pulsar
radiation (e.g.~\cite{vsrr03,hkwe03}). Key tests of all emission
mechanisms and the role of the magnetosphere therefore come from
determining the instantaneous time duration and evolution, the
frequency bandwidth and the polarisation properties of these
phenomena. However, our understanding of how this phenomenology
reflects the underlying physics has been restricted because radio
emission from pulsars is weak and presently there are only a limited
number of sources in which single pulses can be studied
(e.g.~\cite{khk+01}). The unprecedented leap forward in sensitivity
combined with sky coverage provided by the SKA will 
revolutionise the study of single pulses.  To complete this study,
instantaneous coverage of a very wide band (e.g. 0.1 to 5 GHz or higher)
is desireable so that individual pulses can be tracked. This specification
surpasses the strawman 20\% bandwidth specification for the SKA but
would allow new approaches to the understanding of pulsar magnetospheres.

{\em Relationship to Magnetars and Other Objects:} The SKA will also
provide the sensitivity to test whether ``radio-quiet'' objects like
magnetars are in fact very weak radio emitters. The pair cascade
processes that seem to be relevant to create the radio emitting plasma
may be prevented if the magnetic field strength at the polar cap
exceeds the {\em critical magnetic field} 
\begin{equation}
\label{equ:bcrit}
B_{q}=m_e^2 c^3/e\hbar=4.4\times 10^{13} \;\mbox{\rm G}.
\end{equation}
For magnetic fields of such strengths, other processes may compete
with magnetic photon splitting, changing the opacity of the gap
region.  Such arguments have been used to explain the absence of radio
emission from magnetars \cite{bh98}, but these models have been
challenged by the discovery of high magnetic field radio pulsars
\cite{ckl+00,msk+03}. The SKA can both establish whether magnetars
are weak radio emitters as well as discover more radio-loud
pulsars with spin-parameters similar to those of magnetars, so that
the question about the relationship between pulsars and these
prominent high-energy sources can be settled.  As the Galactic 
menagerie of pulsars grows, additional classes of objects may also
be identified, such as young NSs with anemic magnetic fields  that appear
to be lacking on the left side of the $P-\Pdot$ diagram. 

{\em The Radio-High-Energy Connection:} Regularly pulsed high-energy
emission is known for an increasing sample of radio pulsars
\cite{tho00b}. As for the radio emission, the origin of the
high-energy emission is still a matter of significant debate.  Likely
emission processes for the observed non-thermal optical, X-ray and
$\gamma$-ray emission are synchrotron emission, curvature emission and
inverse Compton scattering.  The location of this emission is thought
to be either in the polar cap region (e.g.~\cite{har96}) or in the
outer gaps (e.g.~\cite{chr86a}). In both families of models, the
high-energy emission beam appears to be much wider than the radio
beam. With upcoming missions like GLAST, the high-energy emission of
pulsars will be much better understood by the time observations
with the SKA can be made. Then, however, the SKA can be used to probe
the relationship between high-energy and radio emission, so that a
coherent model of the active magnetosphere can be developed.  Current
results suggest that the high energy emission may indeed be related to
some of the radio emission properties (e.g.~\cite{rj01}).

A radio-high-energy link is provided by the phenomenon of
giant radio pulses. All pulsars with detected giant-pulse emission are
also detected at high energies \cite{jr03}. Another common feature
these pulsars share is the high magnetic field strength at their light
cylinders \cite{cstt96}. Interpretation of this correlation is
unclear at present.  Giant pulses often occur misplaced from the
normal radio profile but appear to be aligned with X-ray and/or
$\gamma$-ray emission (e.g.~\cite{rj01}).  This suggests a common
origin of the giant pulses and high-energy emission and indicates that
some observed radio emission could be in fact a by-product of the
high-energy radiation process.  This could explain highly unusual high
radio-frequency components, seen to emerge at atypical pulse phases in
the Crab profile at a few GHz \cite{mh96}. Observations also suggest
that optical pulses of the Crab pulsars occurring simultaneously with
radio giant pulses appear to be somewhat brighter than other optical
pulses
\cite{sso+03}.

With its sensitivity and monitoring capabilities, the SKA can detect
giant pulses simultaneously at many radio frequencies, allowing us to
determine the spectrum and their relationship to the high energy
emission.  This observational mode is identical to the requirements
for the study of the ``Dynamic Radio Sky'' (see second contribution by
Cordes, this volume), as giant pulse emission is the prototype for
transient coherent emission.  Due to their strength, giant pulses can
be detected from sources as far away as the Virgo cluster and can be
used to directly target young NS in the Milky Way and nearby galaxies,
as discussed in \S\ref{sec:psr.extragal_census}.

\subsection{Resolving Pulsar Magnetospheres}

The SKA can provide the means to actually resolve the pulsar
magnetosphere.  
At a typical distance (1 kpc), the light cylinder radius has an
angular scale $\theta_{\rm LC} \approx 0.32 P D_{\rm kpc}^{-1} \, {\rm
\mu arc sec}$.  Actual radio emission altitudes are much smaller than
$r_{\rm LC}$ and light-travel-time arguments suggest emission sizes of
2-ns duration nano-Giant pulses \cite{hkwe03} about the size of a
beach ball, though relativistic motion toward the observer enlarges
this by a factor $\gamma\sim 10^3$.  Conventional interferometry has
no hope of resolving the relevant scales.  However, interstellar
scintillations (ISS) can probe pulsar emission regions.  
Just as stars twinkle while planets do not because
the critical angular size for atmospheric optical scintillation is
about 1 arcsec, the critical angular size for ISS $\approx 10^{-7}$
arc sec.  Through appropriate measurements of the diffraction pattern
caused by interstellar scattering, the phenomenon may be used
to constrain emission region sizes and locations.  In particular, the
dynamic spectrum $I(\nu, t)$ shows structure in time and frequency
that may differ between different pulse components.  While this method
has been applied to a few pulsars, the number of objects is severely
limited by sensitivity: the pulsar's flux must be detectable in a
narrow frequency channel ($\sim 1$ kHz) in a short time ($\sim 10$
sec).  Also, because the scintillation characteristic bandwidth and
time scale are strongly frequency dependent, flexibility in choosing
the observation frequency and spectrometer resolutions is needed along
with large sensitivity.  As with single-pulse studies, very wide
instantaneous frequency coverage is important for following the
frequency evolution  of ISS and intrinsic effects.  
With the SKA, similar observations will be
possible for a large number of pulsars, therefore contributing to the
final solution of the pulsar problem. Further discussion on resolving
magnetospheres may be found in ``The Microarcsecond Sky and
Cosmic Turbulence,'' (Lazio et al., this volume).

%%%%%%%%%%%%%%%%%%%%%%%%%%%%%%%%%%%%%%%%%%%%%%%%%%%%%%%%%%

\subsection{Extrasolar Planets}
\label{sec:psr.planets}

The first extrasolar planetary system found is around the pulsar
PSR~B1257$+$12 \cite{wf92}.  The system consists of (at least) three
planets, planet~A with approximately a lunar mass, planet~B with a
mass of $4.3\pm0.2 \Mearth$, and planet~C with a mass of~3.9
$\pm0.2 \Mearth$ \cite{kw03}.  Since then, a planet has also been
found around the pulsar PSR~B1620$-$26 \cite{srh+03,rifh03}, in the
globular cluster M4.

While planetary systems around main-sequence stars had been long
anticipated and numerous such systems have been found since, pulsar
planetary systems were not expected.  Although it is unlikely that the
number of pulsar planetary systems will ever approach the number of
planetary systems around main-sequence stars, pulsar planetary systems
offer unique insights.  Taken together the two pulsar planetary
systems already indicate that planets can form in a wide variety of
environments.  Already, for instance, the planet around PSR~B1620$-$26
challenges conventional notions about the formation of planets.  It is
thought to have been acquired by the pulsar during a dynamical
exchange within the globular cluster, implying that this planet has
existed for a substantial fraction of the age of the globular cluster
M4.  This is a low-metallicity globular cluster, suggesting that
planets can form in low-metallicity environments.  In contrast,
planets orbiting main-sequence stars near the Sun are found almost
exclusively around stars with solar- or super-solar metallicities
\cite{gon97,sim01}, which has led to the belief that only stars with
large metal contents can host planets.  Similarly, the presence of
terrestrial mass planets around PSR~B1257$+$12 suggests that
terrestrial planets will be widespread, a prediction to be tested by
future space missions such as Kepler \cite{kbl+98} and the
Terrestrial Planet Finder \cite{bei98}.

In addition to probing the details of planetary formation, 
identification of additional planetary systems around pulsars could
provide valuable clues about pulsar formation and evolution and probe
similarities and differences between planetary systems around pulsars
and main-sequence stars.

\begin{figure}[h]
\includegraphics[width=6cm,angle=-90]{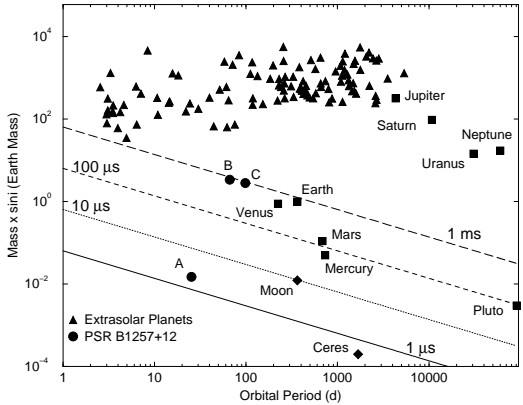}
\vspace{-0.7cm}
\caption{\footnotesize
\label{fig:psr.planets} Planetary companions that can be detected
with pulsar timing observations for different values of timing
precision.  With the SKA a routine timing precision of $\lesssim1\mu$s
is anticipated for millisecond pulsars. Currently detected planets
using optical methods are shown for comparison.  }
\end{figure}

The existence of planets around a pulsar is ascertained from the times
of arrival (TOA) of pulses over a multi-year time span.  The pulsar's
reflex motion in its orbit about the system's barycentre causes the
pulse TOAs to be delayed or advanced relative to what one would
predict for an isolated pulsar.  Thus, finding additional pulsar
planetary systems will require long, high-precision timing
observations. Figure~\ref{fig:psr.planets}, however, demonstrates 
that asteroid-sized objects and smaller can be found with SKA
timing.

%%%%%%%%%%%%%%%%%%%%%%%%%%%%%%%%%%%%%%%%%%%%%%%%%%%%%%%%%%

\subsection{Mapping the Milky Way}
\label{sec:psr.ism}

The goal of a Galactic pulsar census is to detect a large fraction
of the roughly $2 \times 10^4$ pulsars that are beamed toward Earth, 
or about an order of magnitude larger number of pulsars
from current and near-term pulsar surveys.  This census will
necessarily determine the DMs for all of the pulsars as part of the
discovery process,  along with
estimates for the scattering and rotation measures (SM, RM) of a sizable
fraction of the objects in follow-up observations.  We expect
that the SKA will enable a large increase in the number of pulsar
parallaxes, as discussed below. 

The number of sampled lines of sight 
will be sufficiently large and dense on the
sky so as to permit qualitatively new models of the Galaxy to be
developed.  The current ``state of the art'' model for the Galactic
distribution of the warm ionized medium is the NE2001 model
\cite{cl02a,cl02b}.  This model describes the Galaxy largely in terms of
a small number of smooth, large-scale components.  However, it also
incorporates a number of electron density enhancements (``clumps'')
and voids along the line of sight to a limited number of objects
(primarily pulsars).  These clumps and voids are required because the
various observables for these objects (typically pulsar DMs) cannot be
reproduced by only large-scale components.

We envision that, with the Galactic pulsar census provided by the
\hbox{SKA}, it will be possible to trace the Galaxy's structure, or at
least its spiral structure, in a self-consistent manner, rather than
by imposing it as has been done for the NE2001 and previous models.
Parallaxes will be determined for large numbers of pulsars,
providing crucial DM-independent distance measurements by which to
calibrate the local interstellar medium.  Parallaxes will be measureable
for some pulsars out to $\sim 10$ kpc distances.  For many pulsars,
HI absorption can be measured, yielding distance constraints and information
on fine structure in the neutral ISM.
At larger distances, there
should be several pulsars per degree along the Galactic plane.  Their
lines of sight will have a reasonable probability of intersecting an
H\,\textsc{ii} region even over Galactic distances ($\sim 10$~kpc).
The methodology for constructing the model would be to insert clumps
of increased electron density, representing H\,\textsc{ii} regions,
along the line of sight to pulsars.  The clumps would be inserted in a
parsimonious manner so as to minimize the difference between observed
and predicted DMs.  Combined with the parallax measurements and
constraints obtained from scattering measurements of pulsars and
extragalatic sources, it may be possible to identify patterns in the
locations of clumps that would then reveal the spiral arms.
Additional discussion of such modelling is in ``The
Microarcsecond Sky and Cosmic Turbulence,'' (Lazio et al., this volume).

%%%%%%%%%%%%%%%%%%%%%%%%%%%%%%%%%%%%%%%%%%%%%%%%%%%%%%%%%%

\subsection{Mapping the Galactic Center}
\label{sec:psr.mapcenter}

As is the case for the Galaxy as a whole, detecting pulsars in the GC
will allow the magnetoionic medium there to be mapped.  However,
several novel
aspects of the GC result in the potential for additional information
to be obtained.  Firstly, the geometrical weighting
factor for scattering, $W_\tau$, 
can change by large amounts for pulsars at different
distances along the line of sight.  For instance, two pulsars in the GC
separated by only 50~pc along the line of sight ($< 1$\% of the total
distance to the GC) may have pulse broadening times that differ by
25\%.

Secondly, gas motions in the GC can be large, in excess
of~100~km~s${}^{-1}$ \cite{sbv89}.  In contrast to typical
observations through the Galactic disk, for which pulsar velocities
dominate, in the Galactic center, pulsar and gas velocities may be
similar.  If the pulsar and gas velocity vectors are anti-aligned,
changes in pulsar DMs or scattering parameters could happen on much
shorter time scales than are obtained in the Galactic disk.  In 1~yr
the line of sight through a gas cloud could sweep out of order
100~\hbox{AU}, allowing the AU-scale structure of gas clouds in the GC
to be probed.

The distribution of pulse periods that can be detected toward the GC
clearly will be affected by scattering (c.f. Figure~\ref{fig:psr.ngcdetect}).  
Nonetheless, a careful
accounting for selection effects may allow the pulse period
distribution to be used to estimate the past star formation rate in
the GC \cite{lbdh93}.  Even modest constraints on the past star
formation rate would be valuable for assessing the extent to which the
Galaxy's nucleus has undergone episodes of star bursts and their
magnitude.

Finally, one aspect of the SKA's Key Science Project on strong-field
tests of gravity is to map out the space-time around a black hole.  At
a crude level, the distribution of period derivatives of pulsars
around the black hole  could be used to
map out the gravitational potential on the large scales around the
center.  More exciting is the possibility that pulsars will be found
sufficiently close to Sgr~A${}^*$ that strong-field gravitational
effects, like strong micro-lensing or large time delays, will be
measurable \cite{pt79,wgs96,pl03}.  In order to exploit this
possibility for GC pulsars, the line of sight to the pulsar must pass
within a few Einstein radii of Sgr~A${}^*$, or within roughly
$0.\!\!^{\prime\prime}1$.  Consequently, only pulsars within the
central stellar cluster, or those behind the GC with a favourable
projection, will be useful for such strong-field tests.

%%%%%%%%%%%%%%%%%%%%%%%%%%%%%%%%%%%%%%%%%%%%%%%%%%%%%%%%%%

\subsection{Mapping Globular Clusters}
\label{sec:psr.mapcluster}

The dynamical and stellar evolutionary histories of globular clusters
are inexorably linked through the initial binary fraction. The
discovery of tens to hundreds of pulsars in these clusters by the SKA,
combined with the precise timing it will afford, provides an excellent
tool to study these histories. As globular clusters are also thought
to play a vital role in galaxy evolution, an improved understanding of
their formation and history is of importance. Recent work on 47
Tucanae (e.g.~\cite{fck+03}) and NGC 6752 (e.g.~\cite{fps+03}) has
highlighted this potential, in particular when the knowledge
obtained at radio frequencies is combined with results from
X-ray observations (e.g.~\cite{gch+02}).

Using the measured period derivatives of millisecond pulsars it is
possible to measure the line-of-sight acceleration imparted to the
pulsars, as this is usually dominated by the gravitational field of
the globular cluster. When combined with accurate positions and,
possible with the accuracy of SKA timing, accurate parallaxes, this
will lead to the best possible determination of the gravitational
field. Via the mass density distribution and optical observations, the
population of non-optically active cluster members, in particular
massive black holes, can be derived.  The resultant model of the
gravitational potential of the clusters can then be used to determine
the "true" period derivatives of the pulsars and thus their spin-down
energies, magnetic field strengths and ages and also correct any
measured orbital period derivatives.

The more accurate measurement of the gravitational potentials in these
systems will also allow improved modelling of the influence of
binaries on the collapse or otherwise, as well as determinations of how
many pulsars have potentially escaped the cluster and thus the initial
binary fractions. It will also determine what fraction of NSs
have escaped these systems and in turn how many of the NSs
in the Galaxy originated in globular clusters.

The stars in the globular clusters in our Galaxy span a large range of
metallicities and a complete census of pulsars in a large number of
globular clusters will therefore provide vital input in determining
the stellar evolutionary history of clusters and in particular how the
massive end of the initial-mass function depends on
metallicity. Differences in the measured dispersion measures of
pulsars in globular clusters combined either with models of the
potential, or parallaxes, can be used to probe the gas content of the
clusters (\cite{fkl+01}). The gas, originating from the winds of evolved
stars, is expected to be stripped from the cluster, each time the
cluster passes through the Galactic plane every $\sim 10^8$ years
(\cite{spe91}). The amount of gas determined for various clusters
may therefore provide information about the dynamics of the cluster
as a whole.

The study of cluster dynamics is supported by the determination of
the proper motions of the large samples of pulsars. This will allow
accurate determinations of the proper motion of the globular clusters
themselves and thus, when combined with optical proper motions, will
provide excellent ties between radio and optical reference
frames. 

%%%%%%%%%%%%%%%%%%%%%%%%%%%%%%%%%%%%%%%%%%%%%%%%%%%%%%%%%%

\subsection{Pulsar Demographics and Core-Collapse Physics}
\label{sec:psr.population}
\label{sec:psr.supernovae}

The currently known $\sim 1700$ pulsars represent just the `tip of the
iceberg' of a much larger population of perhaps $10^5-10^{6}$ active
pulsars in the Galaxy. The current sample, biased by selection
effects and sensitivity limitations, does not allow answers to a
number of important questions that include: 
\begin{itemize}
\itemsep -3pt
\item How many pulsars are in the Galaxy, what is the birth rate, 
and is there evidence for an increased birth rate in the past?
\item Are neutron stars formed preferentially in spiral arms?  
\item How many pulsars are in globular clusters?
\item Do the magnetic fields of isolated neutron stars decay?
\item What are the minimum and maximum spin periods for radio pulsars?
\item What is the relationship between core collapse and neutron star birth properties?
\item How are isolated millisecond pulsars produced?
\end{itemize}
The properties of the population of pulsars
discovered in the Galactic census (\S\ref{sec:psr.galactic_census}),
combined with the statistics of the discovered
globular cluster (\S\ref{sec:psr.cluster_census}) and
extragalactic pulsars (\S\ref{sec:psr.extragal_census}), will unlock
the answers to these questions.

The outcome of the Galactic Census and future high-energy studies
will establish the relationship between 
the different manifestations of NSs, i.e.~active
radio pulsars, ``radio-quiet'' spin-powered X-ray pulsars, and
Soft-Gamma-Ray Repeaters and Anomalous X-ray Pulsars known
as magnetars (see also \S\ref{sec:psr.relphysics}).

Most (if not all) NSs derive from Type II, core-collapse
supernovae.  Evidence supports the idea that the collapse and ensuring
explosions are not symmetric, thus imparting a translational momentum
impulse or ``kick'' and, perhaps, an angular momentum kick that
determines the initial spin state of the NS.  The large translational
motions of pulsars, 
$\langle V \rangle \sim 500$ km s$^{-1}$ 
represents firm testimony to this picture, with corroboration
from X-ray jets that are aligned with the spin axes in the Crab and
Vela pulsars and also with the proper motion vectors.  While collapse
kicks seem to underly large pulsar velocities, the detailed mechanism
for producing kicks is not yet identified \cite{lcc01}.  Candidate
rocket effects include neutrino emission and mass ejection during NS
formation, which may be altered by the magnetic field strength that
most likely grows during the collapse.  A slower-acting
electromagnetic rocket (the Harrison-Tademaru effect) may occur if NS
are born spinning fast ($P_{\rm initial} \sim 1$ ms) and if the
magnetic field is off axis.  In addition it is not known if the
velocity distribution has been sampled well enough in the
high-velocity component to determine the maximal space velocity and
thus the degree of asymmetry.

With the SKA, comprehensive surveys for pulsars with followup
astrometry (from interferometry) and polarization measurements will
allow a much better understanding of the empirical properties of the
pulsar population with regard to kicks and presumably to
identification of the processes that underly them. Information will
come from determining parallaxes and full 3-D velocities by using the
timing and VLBI capabilities of the SKA. Whilst it is currently
impossible to determine radial velocities for pulsars (unless in orbit
with an optically detectable companion), this will become possible
with the SKA for at least a number of pulsars since the affordable
timing precision will allow us to detect motion on the sky that is
non-linear in time because it results from the three-dimensional spatial
motion of the pulsar projected onto the celestial sphere.

%%%%%%%%%%%%%%%%%%%%%%%%%%%%%%%%%%%%%%%%%%%%%%%%%%%%%%%%%%
%%%%%%%%%%%%%%%%%%%%%%%%%%%%%%%%%%%%%%%%%%%%%%%%%%%%%%%%%%

\section{Technical Requirements}

The key observations to achieve the outlined science goals
can be summarized as
\begin{itemize}
\itemsep -3pt
\item Full Galactic and extragalactic censuses of the pulsar
  population by sensitive searches of the whole sky, the
  Galactic plane, and specific targets such as the
  Galactic Center, Globular Clusters or external galaxies.
\item Regular multi-frequency timing observations 
  of the $\gtrsim 10^4$ discovered pulsars closely spaced in 
  time in about bi-weekly intervals.
\item Targeted observations and dedicated studies
of individual objects and/or areas of special interest.
\item VLBI observations of a large number of sources to determine
  their astrometric parameters.
\end{itemize}
We discuss the technical requirements for these observing programmes
in more detail in the following.

\subsection{Search Observations}
\label{sec:psr.searches}

Yields of pulsar searches need to consider propagation effects and binary
motion in addition to distributions of spin periods, luminosities and
distances of target populations. For both single-pulse and periodicity
searches, algorithms approximate matched filtering procedures on the signal.

The main requirements for pulsar searches are
\begin{enumerate}
\itemsep -3pt
\item A configuration that allows a significant fraction
of the total SKA's collecting area to be used in blind surveys 
that sample a wide field (e.g. at least 1 deg$^2$ at 1 GHz).
\item Frequency coverage that allows optimal detection of pulsars in
periodicity and single-pulse searches for most 
Galactic and extralagalactic lines of sight
(e.g. 0.5 to 5 GHz) and also for the highly scattered Galactic Center
region, which requires 9 to 15 GHz. 
\item Wide bandwidth with adequate channelization to allow dedispersion.
Dispersive smearing across a channel  is
\be
\Delta t_{\rm DM} = \frac{8.3\,\mu s \DM \Delta \nu_{\rm ch}}{\nu^3},
\ee
for $\Delta\nu_{\rm ch}$ and $\nu$ in MHz and GHz, respectively.  
\DM\ ranges
up to 1200 pc cm$^{-3}$ in the known sample and 
values as large as several thousands are expected for some lines of sight.
Assuming 20\% fractional bandwidth (an underestimate at 1-2 GHz frequencies),
the number of channels needed in post-detection dedispersion is
\be
N_{\rm ch} \approx 
	\frac{1600}{\nu^2} 
	\left( \frac{\DM_{\rm max}}{10^3\, {\rm pc \, cm^{-3}}} \right).
\ee
At low frequencies or high \DM, dedispersion is best done coherently,
requiring access to the pre-detection baseband signal.  For coherent
dedispersion, the only extrinsic smearing effects are from interstellar
scattering and instrumentation.  The signal processing burden is quite
large for coherent dedispersion, however, and will  require 
state-of-the-art processors to achieve desired bandwidths.

\item Time resolution sufficient to sample the shortest pulse widths.
\item Flexibility of the correlator/beam-former to provide the above
sampling.
\end{enumerate}
We quantify the collecting area usable in pulsar
surveys as the fraction $f_c$ in a ``core'' array with which
full-FOV sampling can be performed.  In some designs, the core
array would correspond to those antennas whose signals can be directly
correlated (as opposed to being combined in a beamformer first).   
The system-equivalent flux density for the core array is
$\Ssys/f_c = 2k\Tsys / f_c \Ae = 0.14 f_c^{-1}$ Jy  using the nominal
SKA specification of $2\times 10^4$ m$^2$ K$^{-1}$.

\begin{figure}
\includegraphics[width=7.5cm]{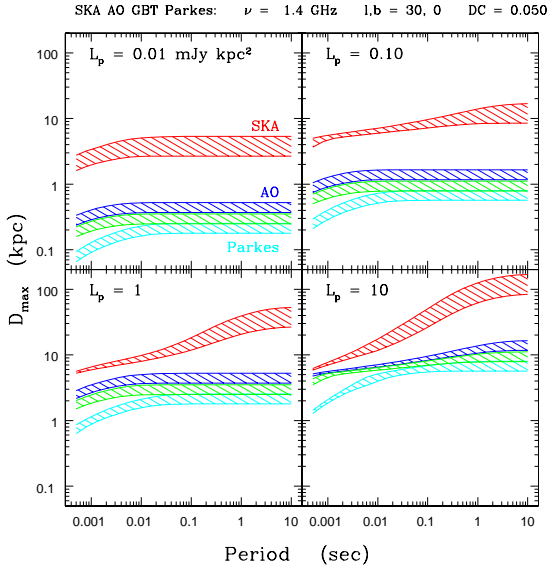}
\caption{\footnotesize
Plot of $\Dmax$ vs.\ $P$ for surveys at~1.4~GHz using the Parkes 64-m antenna,
Arecibo, and the SKA, as labelled, and the Green Bank Telescope (unlabelled). 
The four
panels correspond to different pseudo-luminosities~$L_{\rm p}$.  
In each panel a band is shown for each telescope, the top corresponding
to full gain and the bottom to partial gain.  For the \hbox{SKA}, the partial
gain is due to only $f_c = 0.25$ of the array being available for pulsar
searches.  For the other telescopes, the bottom band boundary is for
0.5 of the peak gain.  Parameters used for the SKA, Arecibo, GBT and
Parkes, respectively, are as follows.  
Bandwidth: 0.4, 0.3, 0.4 and 0.3 GHz;  
Integration time: 3600, 300, 900 and 2100 sec; 
Number of channels: 1024, 1024, 1024 and 96;
Time constant: 64, 64, 64 and 250 $\mu$s;
SEFD: 0.14, 3.6, 16 and 40 Jy.  In all cases, a 10$\sigma$ detection
threshold for the harmonic sum of a Fourier spectrum is assumed.
\break\break
The decline at small periods is due to smearing of pulses from a combination
of receiver time resolution and propagation effects (dispersion and
scattering).   This effect is larger for the Parkes Multibeam survey which
used wide spectrometer channels, thus allowing more dispersive pulse
smearing.  With 1024 channels, dispersion is largely unimportant while
pulse broadening from scattering limits $\Dmax$ for short periods.
periods.
\break\break
The results show that the SKA can detect pulsars at distances 2 to 10
times larger than with Arecibo and can probe to reasonable distances even
for luminosities $L_{\rm p} \lesssim 0.01$ mJy kpc$^2$ that are currently
inaccessible.  For $L_{\rm p} = 10$ mJy kpc$^2$, the SKA can reach 100 kpc
for longer periods.  Given that some pulsars have 
$L_{\rm p} > 10^3$ mJy kpc$^2$ at 1.4 GHz, it is clear the the SKA can
reach distances $> 1$ Mpc in periodicity searches. 
\label{fig:psr.dmax.vs.p}
}
\end{figure}

Plots of SKA performance in pulsar periodicity searches
are shown in Figure~\ref{fig:psr.dmax.vs.p} along with those for existing
telescopes (Arecibo, the Green Bank Telescope, and the Parkes 64m).

{\bf Single-pulse searches:}  Searches for giant pulses can be made in
both blind and targeted searches.  Because they differ from periodicity
searches only in the Fourier analysis part of the latter, the same
requirements on numbers of channels, total bandwidth and overall sensitivity
apply.

%%%%%%%%%%%%%%%%%%%%%%%%%%%%%%%%%%%%%%%%%%%%%%%%%%%%%%%%%%

\subsection{Targeted Observations}
\label{sec:psr.targeted}

In addition to regular timing observations discussed in
\S\ref{sec:psr.timing} and possible VLBI measurements of their astrometric
parameters (\S\ref{sec:psr.vlbi}), many pulsars will also require
dedicated targeted observations to achieve the various science
goals. However, most of the requirements of targeted observations are
identical to those for search and timing measurements.  Indeed, nearly
all aspects of the outlined science programmes have in common that
they require multi-frequency observations. These are necessary to
study the broad-band emission of pulsars, to correct for propagation
effects on the pulsed signal, and for studies of the interstellar
medium itself. It is therefore desirable that
the SKA has constant broad frequency coverage, and at the least
the SKA should be able to change frequency on a regular and rapid
($\sim$ minutes) basis.

For some of science goals, such as the
study of relativistic plasma physics (\S\ref{sec:psr.relphysics}), {\em
simultaneous} multi-frequency observations with full polarization
capabilities are required, often over frequency ranges much greater
than a few GHz.  This simultaneity can be achieved in different ways
with the SKA, depending on the actual design. Assuming that SKA
frontends are made up of a number of different classes of receiver,
each designed to cover a different frequency range, simultaneous
multi-frequency coverage can be achieved through the ability of all of
the receivers in each of the different classes to be operated
simultaneously giving full sensitivity at all
wavelengths. Alternatively, the SKA could form sub-arrays of receivers
from different frequencies up to some limited total number of
receivers.  Whilst this second option still provides the desired
simultaneity, it clearly results in a loss of total sensitivity.

%%%%%%%%%%%%%%%%%%%%%%%%%%%%%%%%%%%%%%%%%%%%%%%%%%%%%%%%%%

\subsection{Timing Measurements}
\label{sec:psr.timing} 

Timing observations require faster sampling of the frequency-time
plane than search observations because the time-tagging of pulses
is desired to high accuracy.   At minimum, dedispersion must be
done as completely as possible, requiring either coherent dedispersion
of baseband signals sampled before detection or post-detection dedispersion
with a large number of channels across the measured bandwidth.  Sample
times needed are $\lesssim 1\,\mu s$ for millisecond pulsars, the objects
of greatest interest.  

{\bf Timing Precision Issues:}
Phenomena that limit time-of-arrival (TOA) precision
include (\cite{cor93})
\begin{enumerate}
\item {\em Radiometer noise}, which yields an rms variation
\be
{\rmstoa}_{,n} &\simeq& \frac{W}{P(S/N)} \nonumber \\ 
	&\simeq& 
	\frac{W\Ssys}{\Save\sqrt{2BT}} \left( \frac{W}{P} \right)^{1/2}, 
\ee
where $S/N$ is the signal to noise ratio of the pulse peak in a sample
pulse profile and $W$ is the pulse width (FWHM, in time units).
\item {\em Pulse phase jitter}, which occurs in the vast majority of
pulsars as drifting sub-pulses and apparently random jumps accompanied
by large amplitude variations having a modulation index 
$m_{\rm I} = $ rms/mean $\sim 1$. Quantifying the rms jitter of a single
pulses as a fraction $f_{\rm j}$ of the pulse width $W$, the rms TOA
variation is
\be
 {\rmstoa}_{,j} \simeq \frac{1}{2}
	\epsilon_i^{-1/2} f_{\rm j} (1+m_{\rm I}^2)^{1/2} 
	W \left( \frac{P}{T} \right)^{1/2},
\ee   
where $\epsilon_i\le 1$ measures the fraction of the pulses in the
observation time $T$ that are statistically independent.  Phase jitter
without intensity modulation ($m_{\rm I} = 0$) produces a TOA error
but the converse is not true.
\item {\em Scattering-induced variations} include the systematic delay
associated with pulse broadening (c.f. \S\ref{sec:psr.center_census}),
which can vary with epoch, and timing variations associated with the
finite number of ``scintles'' (patches of constructive interference)
in the frequency-time plane.  Closely related are DM-variations
associated with large-scale electron-density fluctuations 
(AU to pc; \cite{b+93})
and variations in angle of arrival caused by large-scale gradients.
Together, this collection of effects associated with the dispersion
relation for cold plasma in the ISM yields several contributions with
different frequency scaling, ranging from $\nu^{-2}$ to $\nu^{-4.4}$
\cite{fc90}.
\item {\em Pulse Polarization} can induce TOA errors 
from incorrect gain calibration and from instrumental polarization.
For a Gaussian pulse of width $W$, the induced TOA error from gain variation
is
\be
{\rmstoa}_{\rm ,pol} = \frac{\epsilon\pi_{\rm V} W}{2\sqrt{2\ln 2}}, 
\ee
where $\epsilon$ is the fractional gain error on the circularly
polarized channels ($\delta g/g = \epsilon$) and $\pi_{\rm V}$ is the
degree of circular polarization.  For $\epsilon = \pi_{\rm V} = 0.1$
and $W = $ 1 ms, the error is 4.2 $\mu$s, for example.  If there is no
circular polarization, the error vanishes.  If the receiver
polarization channels are linearly polarized, a similar error ensues.
Instrumental polarization such as cross coupling in the feed 
can 
induce false circular polarization $\pi_{\rm V} \simeq 2\eta^{1/2}
\pi_{\rm L}$, where $\pi_{\rm L}$ is the degree of linear
polarization, which is often quite high in pulsars, and
$\eta$ is the voltage cross coupling coefficient.  Combined with
gain errors, the pseudo-V yields a TOA error as above. To achieve
better than 100 ns timing precision, one must have $\epsilon
\eta^{1/2} \pi_{\rm L} \lesssim 10^{-4} W_{\rm ms}^{-1}$,  
for a pulse width of $W_{\rm ms}$ ms.
If the  gain precision is 1\%,
this requires cross coupling $\eta \lesssim 10^{-4}$, corresponding
to a net polarization purity of $-40$dB.  This may be achieved through
good antenna and feed performance and post-processing calibration.
Similar numbers will hold for linearly polarized feeds, though it is
generally advisable for measurements on pulsars, which are
elliptically polarized, to be made with circularly polarized feeds.
\end{enumerate}

To obtain optimised arrival times, the above scaling laws imply that
observations with the highest sensitivity are needed to minimize the
radiometer noise contribution and thus favour lower frequencies (except
for pulsars with shallow spectra).  Phase jitter, whose strength
(${\rmstoa}_{,j}$) varies greatly from pulsar to pulsar, is nearly
independent of frequency and can be minimised only with longer dwell
times on a pulsar. Phase jitter begins to dominate the contribution
from radiometer noise when S/N $\sim 1$ for a single pulse for the case
where $f_{\rm j} = 1$.  
 For at least one MSP, $f_{\rm j} \ll 1$ \cite{es03}, 
 so that phase jitter may be less important. 
In general, however, high-sensitivity measurements
with the SKA will yield TOAs dominated by phase jitter and systematic
errors from polarization and gain-calibration effects.  Contributions
from propagation effects clearly imply that high frequencies are
favoured. The overall optimal frequency is highly pulsar dependent and
thus requires that the SKA provide a wide range of frequencies,
extending to at least $\sim 5$ GHz.  Moreover, timing measurements at multiple
frequencies allow ISM effects to be partially removed.  

While precision polarization and gain calibrations are required in
order to obtain the best TOAs, the ability to sample the
frequency-time plane with adequate resolution to identify individual
scintillation features is also  required to correct for some of the
perturbations to TOAs.  The effective observing time needed per pulsar
to get a precise TOA measurement will, however, depend on the
individual pulsar, i.e.~as to whether radiometer or pulse-phase jitter
contributions to the TOA uncertainty are dominant.  

{\bf Multiple Beaming and Multiple Fields of View:}
Given the large
number of weak sources to be discovered among the $\sim 20,000$ pulsars
to be timed, the capability to observe multiple pulsars simultaneously,
either through multiple beams within the same FOV or by using multiple
FOVs, is clearly desirable.  Along the Galactic plane we expect 
$\lesssim 100$ sources per 1-deg$^2$ FOV while as many beams
should also be sufficient to sample the denser populated regions.  
However, most 1-deg$^2$ FOVs will contain only a few pulsars,
so that larger FOVs or several independently steerable FOVs
would be beneficial. 

In a worst case analysis, one can estimate from simulations that with
only four independent 1-deg$^2$ FOVs, one needs $\sim 50$ days of
observing time with full-SKA sensitivity to get one TOA measurement
for all discovered pulsars. Assuming that the main contribution to TOA
errors for MSPs is given by radiometer noise, a high S/N TOA
measurement for all $\sim 1000$ MSPs would still require $\sim 10-20$
days with full SKA sensitivity.  Access to only reduced SKA sensitivity
and limited simultaneous multi-frequency coverage
will increase the needed observing time correspondingly while
observations on similar or even more frequent intervals are clearly
necessary to achieve some of the key science goals. However, in
reality, optimisation procedures, such as the choice of lower
frequencies above the Galactic plane or the selection of fewer but
more precise MSPs, will reduce the demand on observing time.
Nevertheless, given the necessity to allow the SKA also to be used for
other key science programmes, a larger number of independently covered
FOVs of the sky appears essential.

%%%%%%%%%%%%%%%%%%%%%%%%%%%%%%%%%%%%%%%%%%%%%%%%%%%%%%%%%%

\subsection{VLBI Observations}
\label{sec:psr.vlbi}

The scientific motivation for obtaining VLBI observations of pulsars
is three-fold.  Firstly, from the requirements of the strong-field
gravity key-science project, it is clear that uncertainties in a
pulsar's distance and motion limit the accuracy to which kinematic
corrections can be applied to observables like the orbital period
derivative and hence to which precision tests of theories of gravity
can be performed (see
\S\ref{sec:psr.grtests}). Secondly, pulsar parallaxes
provide key calibration constraints into models of the Galactic
electron density.  Even a relatively small number of parallaxes can
provide crucial constraints on the distribution of the interstellar
medium near the Sun \cite{tbm+99}. Finally, proper motions are
essential for estimating pulsar velocities, which in turn 
provide constraints on the supernova collapse mechanism
(see \S\ref{sec:psr.supernovae}).

VLBI observations require finding an optimal frequency that balances
astrometric accuracy, pulsar spectra, and atmospheric and ionospheric
 phase
fluctuations.  Pulsar spectra generally favor observations at lower
frequencies ($\lesssim 1$~GHz), but ionospheric phase fluctuations
become significant and the lower resolution attainable (e.g., 5~mas on
terrestrial baselines) vitiate astrometry at these frequencies.
Higher frequencies provide higher astrometric accuracy and reduced
ionospheric phase fluctuations at the cost of a reduced sample size
because of the steep spectra.

We anticipate that the SKA's sensitivity will provide a broad
optimization for VLBI observations in the approximate frequency range
of~2--8~GHz.  Ionospheric phase fluctuations are reduced at these
frequencies while tropospheric phase fluctuations are not yet severe.
Transcontinental/intercontinental baselines provide resolutions of
order 3~mas, and sub-milliarcsecond astrometric accuracies are being
obtained already \cite{ccv+04}.  A key aspect of the SKA's
sensitivity will be its role in finding
calibrators/phase reference sources close to target pulsars.
Chatterjee et al.~(2004)\nocite{ccv+04} have demonstrated that astrometric
accuracy depends strongly upon the separation between the calibrator
and target source.

If 10\% of the SKA's collecting area is spread over
transcontinental/intercontinental baselines, it would be some 20
times more sensitive than current dedicated VLBI arrays.  This
improved sensitivity not only increases the sample of pulsars
available for astrometric purposes but also increases the density of
background sources to serve as phase reference sources.  Finally, we
emphasize the importance of dedicated VLBI capabilities for pulsar
astrometry.  Ad-hoc VLBI arrays can be constructed today with
collecting areas approaching $10^5$~m${}^2$.  By their very nature,
though, these arrays are difficult to maintain for the multi-year
durations required for astrometric projects.

%%%%%%%%%%%%%%%%%%%%%%%%%%%%%%%%%%%%%%%%%%%%%%%%%%%%%%%%%%

\subsection{Summary}
\label{sec:psr.specsum}

The outlined requirements are summarized in Table~\ref{table:spec}
where we quote the required temporal resolution, sensitivity and
frequency coverage, alongside with comments on the required
configuration, Field-of-View (FOV) sampling and  polarization
capabilities.

\begin{sidewaystable}
%\begin{table*}
\caption{\label{table:spec}
Summary of SKA specifications required for the various parts of the
pulsar science case.}
%\begin{tiny}
\begin{tabular}{|l|c|c|c|c|c|c|}
\hline
\hline
Topic & \multicolumn{5}{c}{Required Specification} & \\
\hline
      & $\delta t$ & $A/T$    & $\nu_{\mathrm{max}}$ & Configuration & 
	FOV Sampling & Polarization \\
        %$N_{\mathrm{beams}}$/(1 deg$^2$ FOV) \\
      & ($\mu$s)   & (m$^2$/K) & (GHz) &&& (output, isolation) \\
\hline
\textbf{Searching} &&&&&&    \\
Galactic Census  &  50   & $f_c$ 20,000 & 2.5 & ``core,'' with large $f_c$ & full & Total \\
Galactic Center  &  50   & $f_c$ 20,000 & 15  & ``core,'' with large $f_c$ & full & Total \\
Extragalactic    &  50   & $f_c$ 20,000 & 1.5 & ``core,'' with large $f_c$ & full & Total \\
Globular Clusers &  50   & $f_c$ 20,000 & 1.5 & ``core,'' with large $f_c$ & full & Total \\
Targeted & 50   & $f_c$ 20,000 & 1.5 & ``core,'' with large $f_c$ & full &
 summed \\
								
                 &&&&&&     \\
\textbf{Timing}  &&&&&&     \\
GR tests, NS-BH binaries,  & $<1$ & 20,000 & 15 & non-critical if phasable & 100 beams/deg$^2$ &
	full Stokes, $-40$dB\\
 \quad MSPs &&&&& &  \\
GR tests, Galactic center & 100    & 20,000 & 15 &non-critical if phasable &   10 beams/deg$^2$ & full Stokes, $-40$dB\\
 \quad NS-BH binaries &&&&& &  \\
                 &&&&&&     \\
								
\textbf{VLBI}    &&&&&&      \\
gated astrometry       & 200   & $>2000$   & 8   &  intercontinental baselines & $\sim 3$ beams & Total \\
(proper motion, parallax) &&&&&& \\

\hline
\end{tabular}
%\end{tiny}
%\end{table*}
\end{sidewaystable}

%%%%%%%%%%%%%%%%%%%%%%%%%%%%%%%%%%%%%%%%%%%%%%%%%%%%%%%%%%

\section{Conclusions}\label{sec:psr.conclude}

Given the richness and diversity of the outlined science case, the
research on pulsars conducted since their discovery appears to be only
the prelude to what will be possible with the Square Kilometer
Array. As for other areas in modern astronomy, the SKA will
revolutionise the field of pulsar astrophysics. The SKA will enable
not only the discovery of most pulsars in the Milky Way but also
allows present-day-survey sensitivities to pulsars in the closest
Galaxies.  With the single-pulse search techniques, it should be
possible to detect giant pulses from pulsars as distant as the Virgo
Cluster. The impressive yield of 10,000 to 20,000 discovered pulsars,
including over 1,000 millisecond pulsars, effectively samples every
possible outcome of the evolution of massive binary stars, thereby
guaranteeing the discovery of very exciting systems. We expect at
least 100 compact relativistic binaries, including the elusive
pulsar-black hole systems. For tests of theories of gravity of general
relativity such a system could well surpass all present and
foreseeable competitors.  It may be possible through pulsar timing to
precisely determine the mass, spin and quadrupole moment of the black
hole (see contribution by Kramer et al., this volume). Such measurements
would go beyond what is currently possible with binary pulsar tests of
general relativity, perhaps even confronting Einstein's famous
``no-hair'' theorem for black holes.  The structure of the Milky Way
and its magnetic field can be studied in far greater detail than
presently possible, particularly with the expected increase in
parallax measurements made possible by the SKA's superior
sensitivity. In addition, studies of the significant numbers of
extragalactic pulsars expected to be detectable by the SKA would allow
measurements of the intergalactic, as opposed to the interstellar,
medium.

%\input references.30aug04.tex

% \\bibitem{} %i, %T, %J, %v,(%y), %p-%P.

\end{document}